\documentclass[twocolumn,showpacs,amssymb,aps]{revtex4}
\usepackage{graphicx}
\usepackage{amsbsy}

\begin{document}

\title{Phonon-mediated electron pairing in graphene}
\author{Yu.\,E.~Lozovik${}^{1,2}$}
\email{lozovik@isan.troitsk.ru}
\author{A.\,A.~Sokolik${}^1$}
\affiliation{${}^1$ Institute of Spectroscopy, Russian Academy of Sciences, 142190 Troitsk, Moscow reg.\\
${}^2$ Moscow Institute of Physics and Technology (State University), 141700 Dolgoprudny, Moscow reg.}%

\begin{abstract}
The possibility of superconducting pairing of electrons in doped graphene due to in-plane and out-of-plane phonons is studied. Quadratic coupling of
electrons with out-of-plane phonons is considered in details, taking into account both deformation potential and bond-stretch contributions. The
order parameter of electron-electron pairing can have different structures due to four-component spinor character of electrons wave function. We
consider $s$-wave pairing, diagonal on conduction and valence bands, but having arbitrary structure with respect to valley degree of freedom. The
sign and magnitude of contribution of each phonon mode to effective electron-electron interaction turns out to depend on both the symmetry of phonon
mode and the structure of the order parameter. Unconventional orbital-spin symmetry of the order parameter is found.
\end{abstract}


\maketitle

\section{Introduction}
Low-energy dynamics of electrons in graphene, a two-dimensional form of carbon, is described by a two-dimensional Dirac-type equation for massless
particles \cite{CastroNeto}. Such unusual electronic properties of graphene offer a possibility to study effectively ultrarelativistic electrons
involved in condensed matter phenomena \cite{Katsnelson1,Katsnelson2}, and particularly in collective electron phenomena \cite{Lozovik1}. In the
present paper, we consider qualitatively and estimate quantitatively Bardeen-Cooper-Schrieffer-like (BCS-like) \cite{BCS} phonon-mediated pairing of
electrons in graphene, taking into account the ultrarelativistic electron dynamics. The analogy exists between superconducting pairing of electrons
in graphene and ``color''-superconducting pairing in dense quark matter \cite{Alford}.

As possible origins of electron pairing in graphene, phonon- and plasmon-mediated mechanisms \cite{Uchoa}, electron correlations
\cite{Black-Schaffer,Honerkamp} and highly anisotropic electron-electron scattering near van Hove singularity \cite{Gonzalez} were proposed.
Moreover, superconductivity can be induced in graphene due to proximity effect near superconducting contacts \cite{Heersche,Beenakker}. We study
electron-electron pairing by in-plane optical phonons, represented by four modes with different symmetries, clearly seen in the Raman spectra of
graphene (see, e.g., \cite{Piscanec,Basko1,Basko2,Gruneis}). Furthermore, we consider the quadratic interaction of graphene electrons with two modes
of out-of-plane (flexural) phonons and study a possibility of electron pairing by these phonons (see also the qualitative study in
\cite{Khveshchenko}). The properties of out-of-plane phonons and their interaction with electrons has a close relation to the formation of ripples in
suspended graphene sheets \cite{Meyer,Fasolino} and the the influence of ripples on electrons via effective gauge field \cite{Kim}. The role of
long-wavelength acoustic out-of-plane phonons in low-temperature transport of electrons in graphene was considered in \cite{Mariani1}.

Electron pairing in graphene, considered in the papers \cite{Ohsaku3,Aleiner,Kopnin,Lozovik3,Lozovik4,Lozovik5} within various models, can
demonstrate various peculiarities, in particular, a multi-band character, when electrons from both conduction and valence bands are involved
coherently into the pairing. Eliashberg multi-band equations for phonon-mediated pairing in graphene were derived and solved in \cite{Lozovik5} with
neglecting details of electron-phonon interaction. Such details are taken into account in the present paper, where generally multi-band electron
pairing is considered and the results for the superconducting gap in the one-band limit are presented.

By means of matrix diagrammatic technique, we demonstrate that the effective interaction, induced by each phonon mode and entering the
Eliashberg-type gap equations for electron-electron pairing, depends essentially on symmetry properties of this mode and on the structure of the
electron Cooper pair condensate with respect to the valley degree of freedom (the analogue of chirality in graphene \cite{Jackiw}). In result, the
phonon mode can produce not only an effective attraction, but even effective repulsion. Estimates of the coupling constants show that the
quadratically-coupled out-of-plane phonons do not cause a pairing with any observable critical temperatures, however the in-plane optical phonons can
lead to the pairing in heavily doped graphene.

The article is organized as follows. In Sec.~2 we formulate the Hamiltonian of graphene electrons, interacting with in-plane optical phonons. In
Sec.~3, we derive from the first principles the Hamiltonian of electrons interaction with out-of-plane phonons. Electron-electron pairing due to
in-plane and out-of-plane phonons is considered in Sec.~4 and Sec.~5 respectively, and Sec.~6 is devoted to discussion and conclusions.

\section{Hamiltonian of electrons and in-plane phonons}
\begin{figure}[t]
\begin{center}
\resizebox{0.85\columnwidth}{!}{\includegraphics{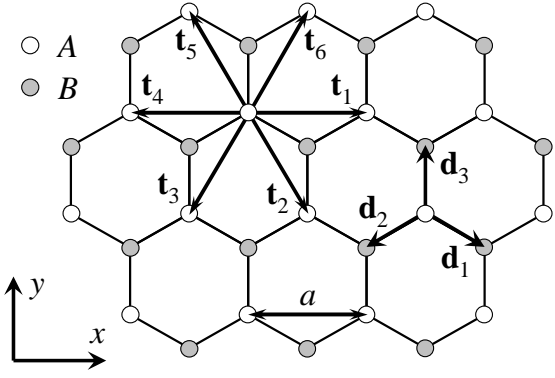}}
\end{center}
\caption{\label{Fig1} Graphene lattice as a combination of two triangular sublattices $A$ and $B$ with the period $a=2.46\,\mbox{\AA}$. The vectors
$\mathbf{d}_k$, ($k=1,2,3$) and $\mathbf{t}_l$ ($l=1,\ldots,6$) connect an atom of the sublattice $A$ with its nearest and next-to-nearest neighbors
respectively.}
\end{figure}

In this paragraph, we consider a Hamiltonian of interacting electrons and in-plane optical phonons in graphene as a base of subsequent consideration
of the pairing. The crystal lattice of graphene consists of two interpenetrating triangle lattices $A$ and $B$ with a period $a=2.46\,\mbox{\AA}$
(Fig.~\ref{Fig1}). We use the Hamiltonian of noninteracting graphene electrons in the tight-binding approximation \cite{CastroNeto}:
\begin{eqnarray}
H_0=-t\sum_{\langle ij\rangle}\left\{a_i^+b_j+b_j^+a_i\right\},\label{H0_1}
\end{eqnarray}
where the sum $\langle ij\rangle$ is taken over pairs of nearest neighbors and $t$ is the hopping amplitude; $a_i$ and $b_j$ are destruction
operators for electrons on the $i$-th and $j$-th sites of sublattices $A$ and $B$. Performing in (\ref{H0_1}) the Fourier transform
$a_i=N^{-1/2}\sum_{\mathbf{p}}e^{i\mathbf{p}\mathbf{r}_i}a_{\mathbf{p}}$, $b_j=N^{-1/2}\sum_{\mathbf{p}}e^{i\mathbf{p}\mathbf{r}_j}b_{\mathbf{p}}$
($N=2S/a^2\sqrt3$ is the number of elementary cells in the crystal and $S$ is the system area), we get
\begin{eqnarray}
H_0=-t\sum_{\mathbf{p}}\left\{a^+_{\mathbf{p}}b_{\mathbf{p}}g_{\mathbf{p}}+b^+_{\mathbf{p}}a_{\mathbf{p}}g_{-\mathbf{p}}\right\},\label{H0_2}
\end{eqnarray}
where
\begin{eqnarray}
g_{\mathbf{p}}=\sum_{k=1}^{3}e^{i\mathbf{p}\mathbf{d}_k},\nonumber
\end{eqnarray}
and the vectors $\mathbf{d}_k$ are shown in Fig.~\ref{Fig1}.

\begin{figure}[t]
\begin{center}
\resizebox{0.6\columnwidth}{!}{\includegraphics{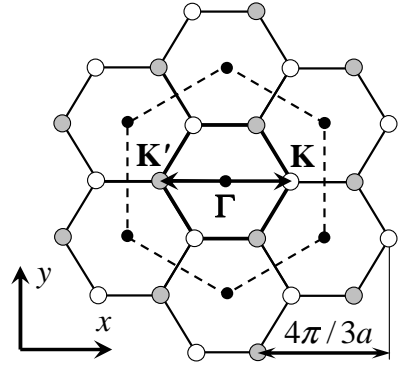}}
\end{center}
\caption{\label{Fig2}The reciprocal lattice of graphene in momentum space. The first and second Brillouin zones are encircled with thick solid and
thin dashed lines respectively. The arrays of equivalent $\Gamma$, $\mathbf{K}$ and $\mathbf{K}'$ points are denoted by circles of different colors.}
\end{figure}

To describe low-energy electron dynamics in graphene, we focus on vicinities of the Dirac points $\mathbf{K}$ and $\mathbf{K}'=-\mathbf{K}$ in
momentum space (Fig.~\ref{Fig2}). Introducing (analogously to \cite{Gusynin}) the four-component spinor operator
$\Psi_{\mathbf{p}}=(a_{\mathbf{K}+\mathbf{p}},b_{\mathbf{K}+\mathbf{p}},b_{\mathbf{K}'+\mathbf{p}},a_{\mathbf{K}'+\mathbf{p}})^T$, the``covariant''
coordinates $p^0=(i/v_\mathrm{F}) (\partial/\partial t)$, $p^{1,2}=p_{x,y}$, $p_\mu=\{p^0,-p^1,-p^2\}$ (here
$v_\mathrm{F}=at\sqrt{3}/2\approx10^6\,\mbox{m/s}$ is the Fermi velocity), and the Dirac gamma-matrices in the Weyl representation
\begin{eqnarray}
\gamma^0=\left(\begin{array}{cc}0&I\\I&0\end{array}\right),\quad\boldsymbol\gamma=\left(\begin{array}{cc}0&-\boldsymbol\sigma\\
\boldsymbol\sigma&0\end{array}\right),\label{gamma_matr}
\end{eqnarray}
where $\boldsymbol\sigma=\{\sigma_x,\sigma_y,\sigma_z\}$, we reduce (\ref{H0_2}) to the form:
\begin{eqnarray}
H_0=v_\mathrm{F}\sum_{\mathbf{p}}\overline\Psi_{\mathbf{p}}\boldsymbol\gamma\mathbf{p}\Psi_{\mathbf{p}},\label{H0_4}
\end{eqnarray}
where $\overline\Psi_{\mathbf{p}}=\Psi^+_{\mathbf{p}}\gamma^0$. The operators $\Psi_{\mathbf{p}}$ in the Heisenberg representation obey the
Dirac-type equation:
\begin{eqnarray}
p_\mu\gamma^\mu\Psi_{\mathbf{p}}=0,\quad\mu=0,1,2.\label{Dirac}
\end{eqnarray}

The most general form of the Hamiltonian, describing the linear coupling of graphene electrons to in-plane phonons, reads
\begin{eqnarray}
H^\mathrm{(lin)}_\mathrm{el-ph}=\frac1{\sqrt{S}}\sum_{\mathbf{p}\mathbf{q}\mu}
g^{(\mu)}_{\mathbf{p}\mathbf{q}}\overline\Psi_{\mathbf{p}+\mathbf{q}}\Gamma_\mu\Psi_{\mathbf{p}}\Phi_{\mathbf{q}\mu},\label{H_int_1}
\end{eqnarray}
where $\mu$ enumerates the phonon modes, and $g^{(\mu)}_{\mathbf{p}\mathbf{q}}$ and $\Gamma_\mu$ are corresponding coupling amplitudes and
interaction vertices; $\Phi_{\mathbf{q}\mu}=c_{\mathbf{q}\mu}+c^+_{-\mathbf{q}\mu}$, where $c_{\mathbf{q}\mu}$ is the phonon destruction operator.

The in-plane phonon modes, most strongly coupled to electrons, are represented by $A_1$ and $B_1$ modes (we denote them by $\mu=1$ and 2
respectively) with the momentum $\mathbf{q}=\pm\mathbf{K}$ and energy $\omega_{\mathbf{K}}\approx0.150\,\mbox{eV}$, and by $E_{2x}$ and $E_{2y}$
modes ($\mu=3$ and 4) with $\mathbf{q}=\Gamma$, $\omega_\Gamma\approx0.196\,\mbox{eV}$ \cite{Piscanec,Basko1,Basko2,Gruneis}. The interaction
vertices for these modes
\begin{eqnarray}
\Gamma_1=1,\quad\Gamma_2=i\gamma^5,\quad\Gamma_3=-\gamma^5\gamma^2,\quad\Gamma_4=-\gamma^5\gamma^1\label{vert1}
\end{eqnarray}
reflect their symmetry: scalar ($\mu=1$), pseudoscalar ($\mu=2$) and pseudovector ($\mu=3,4$). The coupling constants
$g^{(\mu)}_{\mathbf{p}\mathbf{q}}$ are weakly dependent on $\mathbf{p}$, $\mathbf{q}$: $g^{(\mu)}_{\mathbf{p}\mathbf{q}}\equiv g_\mu$, and can be
related to the values $\langle g^2_{\mathbf{K},\Gamma}\rangle_\mathrm{F}$, introduced in \cite{Piscanec}: $g_{1,2}^2=2\sqrt3a^2\langle
g^2_{\mathbf{K}}\rangle_\mathrm{F}$, $g_{3,4}^2=2\sqrt3a^2\langle g^2_\Gamma\rangle_\mathrm{F}$.

\section{Interaction of electrons with out-of-plane phonons}
For out-of-plane phonon modes, we cannot restrict ourselves to phonon momenta close to $\Gamma$ and $\pm\mathbf{K}$, as will be shown below.
Therefore, we need a detailed microscopic model description of them. We start this description from the simplified version of the valence force
Lagrangian \cite{Perebeinos,Kuzminskiy}, taking into account only the bond-bending term:
\begin{eqnarray}
\mathcal{L}=\frac{M}2\sum_i(u_i^A)^2+\frac{M}2\sum_j(u_j^B)^2\nonumber\\-\frac{D}2\sum_i\left[\sum_{j\in
N_1(i)}(u_j^B-u_i^A)\right]^2\nonumber\\-\frac{D}2\sum_j\left[\sum_{i\in N_1(j)}(u_i^A-u_j^B)\right]^2.\label{Flex1}
\end{eqnarray}
Here $u_i^A$ and $u_j^B$ are out-of-plane displacements of carbon atoms from the sublattices $A$ and $B$ respectively, $M$ and $D$ are the carbon
atom mass and the elasticity coefficient; $N_p(i)$ denotes the set of $p$-th order neighbors of the $i$-th atom. The Lagrange equations for atomic
displacements, following from (\ref{Flex1}), are:
\begin{eqnarray}
\left(M\frac{\partial^2}{\partial t^2}+12D\right)u_i^{A,B}-6D\sum_{j\in N_1(i)}u_j^{B,A}\nonumber\\+D\sum_{k\in N_2(i)}u_k^{A,B}=0.\label{Flex2}
\end{eqnarray}

The displacements $u_i^{A,B}$ can be decomposed over out-of-plane phonon modes with definite momenta $q$, covering the second Brillouin zone (see
Fig.~\ref{Fig2}), and branches $\sigma$ ($\sigma$=1 for acoustical branch and 2 for optical one):
\begin{eqnarray}
u_i^{A,B}=\sum_{\mathbf{q}\sigma}\frac{\varepsilon_{\mathbf{q}\sigma}^{A,B}}{\sqrt{6NM\omega_{\mathbf{q}\sigma}}}\,
\Phi_{\mathbf{q}\sigma}e^{i\mathbf{q}\mathbf{r}_i},
\label{Flex3}
\end{eqnarray}
where $\varepsilon_{\mathbf{q}\sigma}^{A,B}$ are polarizations of the phonon modes;
$\Phi_{\mathbf{q}\sigma}=c_{\mathbf{q}\sigma}+c^+_{-\mathbf{q}\sigma}$ is the phonon operator (in Heisenberg representation) and
$\omega_{\mathbf{q}\sigma}$ is its frequency. Substituting (\ref{Flex3}) into (\ref{Flex2}), we obtain the system for determination of
characteristics of phonon modes:
\begin{eqnarray}
\left(\begin{array}{cc}\frac{M}D\omega_{\mathbf{q}\sigma}^2-h_{\mathbf{q}}&6g_{\mathbf{q}}\\
6g^*_{\mathbf{q}}&\frac{M}D\omega_{\mathbf{q}\sigma}^2-h_{\mathbf{q}}\end{array}\right)\left(\begin{array}{c}\varepsilon^A_{\mathbf{q}\sigma}\\
\varepsilon^B_{\mathbf{q}\sigma}\end{array}\right)=0,\label{Flex4}
\end{eqnarray}
where $h_{\mathbf{q}}=12+\sum_{l=1}^6e^{i\mathbf{q}\mathbf{t}_l}$ (see Fig.~\ref{Fig1} for definition of $\mathbf{t}_l$). Solving (\ref{Flex4}), we
find:
\begin{eqnarray}
\omega_{\mathbf{q}\sigma}^2=\frac{D}M\left(h_{\mathbf{q}}-6s_\sigma|g_{\mathbf{q}}|\right),\nonumber\\
\varepsilon_{\mathbf{q}\sigma}^A=s_\sigma\frac{g_{\mathbf{q}}}{\sqrt2|g_{\mathbf{q}}|},\quad\varepsilon_{\mathbf{q}\sigma}^B=\frac1{\sqrt2},\label{Flex5}
\end{eqnarray}
where $s_1=+1$, $s_2=-1$. In principle, the summation over $\mathbf{q}$ in (\ref{Flex3}) within the first Brillouin zone is sufficient, but it is
convenient in our approach to use the second Brillouin zone, since the quantities (\ref{Flex5}) are defined unambiguously there. The long-wavelength
asymptotics of the frequencies (\ref{Flex5}) are: $\omega_{\mathbf{q}1}\approx(a^2/4)\sqrt{D/M}q^2$, $\omega_{\mathbf{q}2}\approx6\sqrt{D/M}$.
Comparing $\omega_{\mathbf{q}2}$ with the experimental data \cite{Wirtz}, we get $\sqrt{D/M}\approx0.019\,\mbox{eV}$.

Two mechanisms provide the main contribution to electron-phonon coupling in graphene: the deformation potential and the change of bond lengths
\cite{Mariani1,Suzuura}. In the simplest approximation, the deformation potential, acting on electron bound to specific carbon atom, originates from
the potentials of the nearest neighbors of this atom, approaching it (or moving away from it), and thus is determined by the sum of lengths of bonds,
connecting this atom with its nearest neighbors. At the same time, the change of bond lengths modulate the hopping integrals in (\ref{H0_2}). In the
limit of small out-of-plane displacements, the change of distance between $i$-th atom of the $A$ sublattice and $j$-th atom of the $B$ sublattice is
approximately $(u_j^B-u_i^A)^2\sqrt3/2a$. Therefore the Hamiltonian of quadratic electron-phonon coupling is:
\begin{eqnarray}
H^\mathrm{(quadr)}_\mathrm{el-ph}=\sum_{\langle ij\rangle}(u_j^B-u_i^A)^2\left\{C_1(a_i^+a_i+b_j^+b_j)\right.\nonumber\\
\left.+C_2(a_i^+b_j+b_j^+a_i)\right\},\label{H_int_2}
\end{eqnarray}
where the multipliers of $C_1$ and $C_2$ correspond to the deformation potential and bond-stretch contributions; $C_2=\gamma\sqrt3/2a$,
$\gamma=\partial t/\partial a\approx6\,\mbox{eV/\AA}$ \cite{Basko2}. Taking the continuum long-wavelength limit of the deformation potential part of
(\ref{H_int_2}) and comparing it with \cite{Mariani1,Suzuura}, we find $C_1=g_1/a^2$, where $g_1\approx20-30\,\mbox{eV}$.

Performing the Fourier transform for electron operators in (\ref{H_int_2}) and using (\ref{Flex3}), we get
\begin{eqnarray}
H^\mathrm{(quadr)}_\mathrm{el-ph}=\sum_{\mathbf{P}\mathbf{P}'}\sum_{\mathbf{q}\sigma\sigma'}
\frac{\Phi_{\mathbf{q}\sigma}\Phi_{\mathbf{q}'\sigma'}}{6NM\sqrt{\omega_{\mathbf{q}\sigma}\omega_{\mathbf{q}'\sigma'}}}\nonumber\\
\times\sum_{L_1L_2}\left\{a^+_{\mathbf{P}'}a^{\vphantom{+}}_{\mathbf{P}}R^{AA}_{L_1L_2}+b^+_{\mathbf{P}'}b^{\vphantom{+}}_{\mathbf{P}}R^{BB}_{L_1L_2}
\right.\nonumber\\
+\left.a^+_{\mathbf{P}'}b^{\vphantom{+}}_{\mathbf{P}}R^{AB}_{L_1L_2}+b^+_{\mathbf{P}'}a^{\vphantom{+}}_{\mathbf{P}}R^{BA}_{L_1L_2}\right\}
\varepsilon^{L_1}_{\mathbf{q}\sigma}\varepsilon^{L_2}_{\mathbf{q}'\sigma'}.\label{H_int_3}
\end{eqnarray}
Here $\mathbf{P}$ and $\mathbf{P}'$ run over the first Brillouin zone, and $\mathbf{q}'=\mathbf{P}'-\mathbf{P}-\mathbf{q}$. For $L_1,L_2=A,B$ we have
introduced the matrices:
\begin{eqnarray}
R^{AA}_{L_1L_2}=C_1\left(\begin{array}{cc}1&-g_{\mathbf{q}'}\\-g_{\mathbf{q}}&g_{\mathbf{P}'-\mathbf{P}}\end{array}\right)_{L_1L_2},\nonumber\\
R^{BB}_{L_1L_2}=C_1\left(\begin{array}{cc}1&-g_{-\mathbf{q}}\\-g_{-\mathbf{q}'}&g_{-\mathbf{P}'+\mathbf{P}}\end{array}\right)_{L_1L_2},\nonumber\\
R^{AB}_{L_1L_2}=C_2\left(\begin{array}{cc}g_{\mathbf{P}}&-g_{\mathbf{P}'-\mathbf{q}}\\
-g_{\mathbf{P}+\mathbf{q}}&g_{\mathbf{P}'}\end{array}\right)_{L_1L_2},\nonumber\\
R^{BA}_{L_1L_2}=C_2\left(\begin{array}{cc}g_{-\mathbf{P}'}&-g_{-\mathbf{P}-\mathbf{q}}\\
-g_{-\mathbf{P}'+\mathbf{q}}&g_{-\mathbf{P}}\end{array}\right)_{L_1L_2}.\nonumber
\end{eqnarray}

Since electrons populate close vicinities of the momentums $\mathbf{K}$ and $\mathbf{K}'$ (Fig.~\ref{Fig2}), we split the sums in (\ref{H_int_3})
over $\mathbf{P}$ and $\mathbf{P}'$ among the valleys. Using the four-component spinor notations, we can rewrite (\ref{H_int_3}) in the form:
\begin{eqnarray}
H^\mathrm{(quadr)}_\mathrm{el-ph}=\frac1S\sum_{\mathbf{p}\mathbf{p}'\mathbf{q}}\sum_{\mathbf{Q}\sigma\sigma'}\overline\Psi_{\mathbf{p}'}
V^{(\mathbf{Q})}_{\mathbf{p}\mathbf{p}'\mathbf{q},\sigma\sigma'}\Psi_{\mathbf{p}}\Phi_{\mathbf{q}\sigma}\Phi_{\mathbf{q}'\sigma'},\label{H_int_4}
\end{eqnarray}
where $\mathbf{q}'=\mathbf{Q}+\mathbf{p}'-\mathbf{p}-\mathbf{q}$, the electron momentums $\mathbf{p}$, $\mathbf{p}'$ are small and measured from the
Dirac points, and the vector $\mathbf{Q}$ takes the values $\Gamma$ and $\pm\mathbf{K}$. We can simplify cumbersome expressions for the vertices,
assuming that they depend slowly on $\mathbf{p}$ and $\mathbf{p}'$. Denoting $V^{(\mathbf{Q})}_{00\mathbf{q},\sigma\sigma'}\equiv
V^{(\mathbf{Q})}_{\mathbf{q}\sigma\sigma'}$ and using (\ref{Flex5}), we get:
\begin{eqnarray}
V^{(\Gamma)}_{\mathbf{q}\sigma\sigma'}=\frac{a^2}{8M\sqrt{\omega_{\mathbf{q}\sigma}\omega_{\mathbf{q}\sigma'}}}\left\{\vphantom{\frac11}
2\delta_{\sigma\sigma'}C_1(3-s_\sigma|g_{\mathbf{q}}|)\gamma^0\right.\nonumber\\+C_2\gamma^5\left[\left(s_\sigma
g_{\mathbf{K}-\mathbf{q}}\phi_{\mathbf{q}}+s_{\sigma'}g_{\mathbf{K}+\mathbf{q}}\phi_{-\mathbf{q}}\right)\frac{\gamma^1+i\gamma^2}2\right.\nonumber\\
\left.\left.+\left(s_\sigma
g_{-\mathbf{K}-\mathbf{q}}\phi_{\mathbf{q}}+s_{\sigma'}g_{-\mathbf{K}+\mathbf{q}}\phi_{-\mathbf{q}}\right)\frac{\gamma^1-i\gamma^2}2\right]\right\},
\nonumber\\
V^{(\mathbf{K})}_{\mathbf{q}\sigma\sigma'}=
\frac{a^2}{4M}\frac{1+\gamma^5}{4\sqrt{\omega_{\mathbf{q}\sigma}\omega_{\mathbf{K}-\mathbf{q},\sigma'}}}\nonumber\\
\times\left\{\vphantom{\frac11}C_1(3-s_\sigma|g_{\mathbf{q}}|-s_{\sigma'}|g_{\mathbf{K}-\mathbf{q}}|)\right.\nonumber\\
\times\gamma^0\left(\frac{\gamma^1+i\gamma^2}2+\frac{\gamma^1-i\gamma^2}2s_\sigma
s_{\sigma'}\phi_{\mathbf{q}}\phi_{\mathbf{K}-\mathbf{q}}\right)\nonumber\\
\left.+C_2(s_\sigma g_{-\mathbf{K}-\mathbf{q}}\phi_{\mathbf{q}}+
s_{\sigma'}g_{\mathbf{K}+\mathbf{q}}\phi_{\mathbf{K}-\mathbf{q}})\vphantom{\frac11}\right\},
\nonumber\\
V^{(-\mathbf{K})}_{\mathbf{q}\sigma\sigma'}=\gamma^0\left[V^{(\mathbf{K})}_{-\mathbf{q}\sigma\sigma'}\right]^+\gamma^0,\label{vert2}
\end{eqnarray}
where $\phi_{\mathbf{q}}=g_{\mathbf{q}}/|g_{\mathbf{q}}|$.

\section{Electron pairing by in-plane phonons}
To describe pairing, we introduce the following set of matrix $(4\times4)$ Green functions in Matsubara representation (similarly to
\cite{Pisarski}):
\begin{eqnarray}
G_{ij}(\mathbf{p},\tau)=-\langle T\Psi^{(i)}_{\mathbf{p}}(\tau)\overline\Psi^{(j)}_{\mathbf{p}}(0)\rangle,\label{G_ij}
\end{eqnarray}
where $\Psi^{(1)}_{\mathbf{p}}(\tau)=\Psi_{\mathbf{p}}(\tau)$, $\Psi^{(2)}_{\mathbf{p}}(\tau)=\Psi_{\mathrm{C}\mathbf{p}}(\tau)\equiv
C\overline\Psi^T_{-\mathbf{p}}$ is the charge-conjugated spinor and $C=i\gamma^2\gamma^0$ is the charge-conjugation matrix. The diagonal elements of
(\ref{G_ij}), $G_{11}$ and $G_{22}$, are the Green functions of particles and ``antiparticles'', while $G_{12}$ and $G_{21}$ describe a Cooper pair
condensate. The Green functions of free particles are: $G_{11}^{(0)}(p)=[\gamma^0(p_0+\mu)-v_\mathrm{F}\boldsymbol\gamma\mathbf{p}]^{-1}$,
$G_{22}^{(0)}(p)=[\gamma^0(p_0-\mu)-v_\mathrm{F}\boldsymbol\gamma\mathbf{p}]^{-1}$, where $p=\{p_0=i\pi T(2n+1),\mathbf{p}\}$ and $\mu$ is the
chemical potential in graphene, measured from the Dirac points.

\begin{figure}[t]
\begin{center}
\resizebox{0.75\columnwidth}{!}{\includegraphics{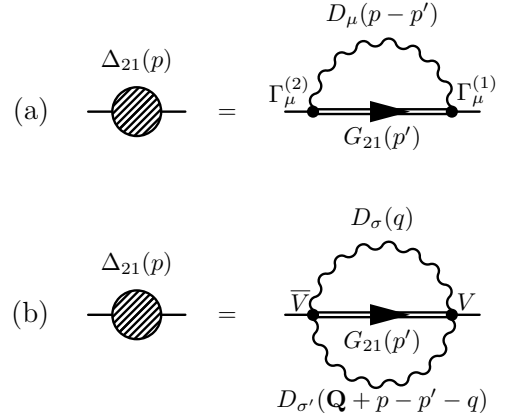}}
\end{center}
\caption{\label{Fig3} Diagrammatic representation of self-consistent gap equations: (a) for linear electron-phonon coupling (\ref{Sc1}), (b) for
quadratic electron-phonon coupling (\ref{Sc5}).}
\end{figure}

Employing the standard diagrammatic technique for the system of graphene electrons with the Hamiltonian (\ref{H0_4}), interacting with in-plane
phonons via (\ref{H_int_1}), we get the following set of Gor'kov-type equations for the Green functions (\ref{G_ij}), describing the pairing in the
mean-field approximation:
\begin{eqnarray}
G_{ij}(p)=G_{ij}^{(0)}(p)-\frac{T}S\sum_{p'k\mu}g_\mu^2D_\mu(p-p')G_{ii}^{(0)}(p)\nonumber\\
\times\Gamma_\mu^{(i)}G_{ik}(p')\Gamma_\mu^{(k)}G_{kj}(p),\label{gor1}
\end{eqnarray}
where $\Gamma^{(1)}_\mu=\Gamma_\mu$, $\Gamma^{(2)}_\mu=C^{-1}\Gamma_\mu^TC$ is the charge-conjugated vertex and
$D_\mu(q)=2\omega_{\mathbf{q}\mu}/(q_0^2-\omega_{\mathbf{q}\mu}^2)$ is the phonon Green function. Introducing the anomalous self-energies
\begin{eqnarray}
\Delta_{ij}(p)=-\frac{T}S\sum_{p'\mu}g_\mu^2D_\mu(p-p')\Gamma^{(i)}_\mu G_{ij}(p')\Gamma^{(j)}_\mu\label{Sc1}
\end{eqnarray}
for $i\neq j$, we rewrite (\ref{gor1}) in the form
\begin{eqnarray}
G_{11}^{-1}=G_{11}^{(0)-1}-\Delta_{12}G_{22}^{(0)}\Delta_{21},\nonumber\\G_{22}^{-1}=G_{22}^{(0)-1}-\Delta_{21}G_{11}^{(0)}\Delta_{12},\nonumber\\
G_{12}=G_{11}^{(0)}\Delta_{12}G_{22},\quad G_{21}=G_{22}^{(0)}\Delta_{21}G_{11}.\label{gor2}
\end{eqnarray}
The diagrammatic representation of (\ref{Sc1}) is shown in Fig.~\ref{Fig3}(a).

Similarly to \cite{Pisarski,Ohsaku1,Ohsaku2,Ohsaku3}, we employ the following method to solve the matrix equations (\ref{Sc1})--(\ref{gor2}): 1) we
assume a certain form of the order parameter $\Delta_{ij}$, which has a definite matrix structure and is parameterized by a set of variables; 2) with
this $\Delta_{ij}$, we solve the first pair of the Gor'kov equations (\ref{gor2}) and find the inverted normal Green functions $G_{ii}^{-1}$; 3)
inverting $G_{ii}^{-1}$, we substitute $G_{ii}$ into the second pair of the equations (\ref{gor2}), finding the anomalous Green functions $G_{ij}$;
4) substituting $G_{ij}$ into (\ref{Sc1}), we get a closed system of equations for variables, parameterizing $\Delta_{ij}$. The matrix structures of
$G_{ij}$, found from (\ref{gor2}), should correspond to that of the initially assumed $\Delta_{ij}$ via (\ref{Sc1}), and generally this occurs only
for the certain forms of the order parameter $\Delta_{ij}$. Possible matrix structures of the order parameter, describing the pairing of relativistic
elementary particles or electrons in graphene in various models, were discussed in
\cite{Pisarski,Ohsaku1,Ohsaku2,Ohsaku3,Lozovik3,Lozovik4,Capelle,Ryu}.

For phonon-mediated electron pairing, we assume the simplest form of the order parameter: we suppose that the pairing is $s$-wave and diagonal with
respect to conduction and valence bands. In the relativistic-like approach to the electron dynamics (\ref{Dirac}), the states of electron in the
valleys $\pm\mathbf{K}$ are the states with the ``chirality'' quantum numbers $\pm1$ respectively. Moreover, the electron states in conduction band
(positive-energy ``particles'') have the equal chirality and helicity, while the electron states in valence band (negative-energy ``antiparticles'')
have the opposite chirality and helicity. The projection operators on the states with definite chirality ($c=\pm1$) and helicity ($h=\pm1$) are
\cite{Pisarski}:
\begin{eqnarray}
\mathcal{P}_{ch}(\hat{\mathbf{p}})=\frac{1+c\gamma^5}2\times\frac{1+h\gamma^5\gamma^0\boldsymbol\gamma\hat{\mathbf{p}}}2,\nonumber
\end{eqnarray}
where $\gamma^5=i\gamma^0\gamma^1\gamma^2\gamma^3$, $\hat{\mathbf{p}}=\mathbf{p}/|\mathbf{p}|$. The following operators make projections on the
conduction ($\alpha=+1$) and valence ($\alpha=-1$) bands:
\begin{eqnarray}
\tilde\mathcal{P}_\alpha(\hat{\mathbf{p}})=\mathcal{P}_{c,\alpha\cdot h}(\hat{\mathbf{p}})+\mathcal{P}_{-c,-\alpha\cdot
h}(\hat{\mathbf{p}})=\frac{1+\alpha\gamma^0\boldsymbol\gamma\hat{\mathbf{p}}}2.\label{proj}
\end{eqnarray}

We assume an arbitrary structure of the order parameter with respect to valley degree of freedom, represented by $(2\times2)$ unitary matrix in the
space of the valleys $\{\mathbf{K},\mathbf{K}'\}$ (the same was supposed in \cite{Aleiner}, where pairing of Zeeman-split electrons and holes in
graphene was studied). The operators $T_1=\gamma^5$, $T_2=\gamma^3\gamma^5$ and $T_3=i\gamma^3$ allows us to construct the expression for
$\Delta_{ij}$ explicitly. These operators perform rotations of the order parameter in the space of valley degree of freedom and, as noted in
\cite{Gusynin}, obey the algebra of the Pauli matrices. Thus, the general form of the order parameter $\Delta_{21}$, corresponding to the
band-diagonal pairing with the arbitrary valley structure, is
\begin{eqnarray}
\Delta_{21}(p)=e^{i\mathbf{v}\mathbf{T}}\left[\Delta_+(p)\tilde\mathcal{P}_+(\hat{\mathbf{p}})+
\Delta_-(p)\tilde\mathcal{P}_-(\hat{\mathbf{p}})\right],\label{Delta_21}
\end{eqnarray}
where $\Delta_\pm(p)$ are the gaps in conduction and valence bands (see \cite{Lozovik3,Lozovik4}), and the three-dimensional vector $\mathbf{v}$
parameterizes the valley structure of the order parameter.

Substituting (\ref{Delta_21}) into (\ref{gor2}) and taking into account, that $\Delta_{12}=\gamma^0\Delta_{21}^+\gamma^0$, we find:
\begin{eqnarray}
G_{11}(p)=\sum_{\alpha=\pm}\frac{p_0+\xi_\alpha(\mathbf{p})}{p_0^2-E_\alpha^2(p)}\,
\tilde\mathcal{P}_\alpha(\hat{\mathbf{p}})\gamma^0,\nonumber\\
G_{21}(p)=\sum_{\alpha=\pm}\frac{\Delta_\alpha(p)}{p_0^2-E_\alpha^2(p)}\,
\gamma^0e^{i\mathbf{v}\mathbf{T}}\tilde\mathcal{P}_\alpha(\hat{\mathbf{p}})\gamma^0,
\label{G_21}
\end{eqnarray}
where $\xi_\pm(\mathbf{p})=\pm v_\mathrm{F}|\mathbf{p}|-\mu$ and $E_\pm(p)=\sqrt{\xi^2_\pm(\mathbf{p})+\Delta^2_\pm(p)}$ are the excitation energies
for bare electrons and Bogolyubov quasiparticles in conduction and valence bands. Using (\ref{Delta_21}) and (\ref{G_21}) in (\ref{Sc1}), then
multiplying the both parts of (\ref{Sc1}) by $\mathcal{P}_\alpha(\hat{\mathbf{p}})e^{-i\mathbf{v}\mathbf{T}}$ from the left and taking a trace, we
derive the system of self-consistent equations for two gaps $\Delta_\pm(p)$:
\begin{eqnarray}
\Delta_\alpha(p)=\frac{T}S\sum_{p'\mu}\sum_{\beta=\pm}g_\mu^2D_\mu(p-p')\frac{\Delta_\beta(p')}{p_0'^2-E^2_\beta(p')}\nonumber\\
\times R^\mu_{\alpha\beta}(\hat{\mathbf{p}},\hat{\mathbf{p}}';\mathbf{v}),\label{Sc2}
\end{eqnarray}
where the following \emph{angular factors} are introduced:
\begin{eqnarray}
R^\mu_{\alpha\beta}(\hat{\mathbf{p}},\hat{\mathbf{p}}';\mathbf{v})=\nonumber\\-
\frac12\mathrm{Sp}\left[\tilde\mathcal{P}_\alpha(\hat{\mathbf{p}})e^{-i\mathbf{v}\mathbf{T}}
\Gamma_\mu^{(2)}\gamma^0e^{i\mathbf{v}\mathbf{T}}\tilde\mathcal{P}_\beta(\hat{\mathbf{p}}')\gamma^0\Gamma_\mu^{(1)}\right].\label{ang_fact1}
\end{eqnarray}
The contact character of electron interaction with the optical phonons allows to perform angle integration over $\hat{\mathbf{p}}$ and
$\hat{\mathbf{p}}'$ in (\ref{ang_fact1}), which leads to independence of $R^\mu_{\alpha\beta}(\hat{\mathbf{p}},\hat{\mathbf{p}}';\mathbf{v})\equiv
R_\mu(\mathbf{v})$ over $\alpha$ and $\beta$. Calculating the angular factors (\ref{ang_fact1}) using (\ref{vert1}), (\ref{proj}) and summing over
degenerate modes, $R_{\mathbf{K}}=R_1+R_2$ and $R_\Gamma=R_3+R_4$, for $\mathbf{K}$- and $\Gamma$-phonons respectively, we get:
\begin{eqnarray}
R_{\mathbf{K}}=-\cos^2v+\hat{\mathbf{v}}_1^2\sin^2v,\nonumber\\
R_\Gamma=-\cos^2v+(-\hat{\mathbf{v}}_1^2+\hat{\mathbf{v}}_2^2+\hat{\mathbf{v}}_3^2)\sin^2v.\label{ang_fact}
\end{eqnarray}

The system of gap equations (\ref{Sc2}) can be rewritten in the form of two-band gap equations \cite{Lozovik3,Lozovik4,Lozovik5}
\begin{eqnarray}
\Delta_\alpha(p)=\frac{T}S\sum_{p'}\sum_{\beta=\pm}V_\mathrm{eff}(p-p';\mathbf{v})\frac{\Delta_\beta(p')}{p_0'^2-E^2_\beta(p')},\label{Sc3}
\end{eqnarray}
but with the \emph{effective} interaction, dependent on $\mathbf{v}$:
\begin{eqnarray}
V_\mathrm{eff}(q;\mathbf{v})=\sum_\mu g_\mu^2D_\mu(q)R_\mu(\mathbf{v}).\label{V_eff}
\end{eqnarray}
The effective interaction (\ref{V_eff}) is a linear combination of interactions due to separate phonon modes with the coefficients $R_\mu$, which can
take values in the range $-1<R_\mu<1$ depending on $\mathbf{v}$. Therefore, the in-plane optical phonons in graphene can induce not only attraction
between electrons ($R_\mu>0$), but even repulsion ($R_\mu<0$).

Consider now the limiting cases of the valley structure of the order parameter. At $\mathbf{v}=0$, the valley part of the order parameter is the unit
$(2\times2)$ matrix in the valley space; in this case, all $R_\mu=-1$, and all the phonon modes lead to effective repulsion. When
$\mathbf{v}=\{\pi/2,0,0\}$, the valley part of the order parameter is $\sigma_z$, and $R_{\mathbf{K}}=1$ (effective attraction), $R_\Gamma=-1$
(repulsion). Finally, at $\mathbf{v}=\{0,\pi/2,0\}$ and $\mathbf{v}=\{0,0,\pi/2\}$, the valley part is $\sigma_x$ and $\sigma_y$ respectively
(valley-off diagonal pairing), and $R_{\mathbf{K}}=0$ (mutual cancelation of contributions from scalar $\mu=1$ and pseudoscalar $\mu=2$ phonon
modes), $R_\Gamma=1$ (attraction).

The maximal value of effective electron-phonon coupling constant can be reached in highly doped graphene due to large density of states at the Fermi
level $\mathcal{N}=\mu/2\pi v_\mathrm{F}^2$. At $\mu>\omega_\mu$ the pairing can be treated as one-band (i.e. involving only the conduction band). In
\cite{Lozovik5}, the equations of the type of (\ref{Sc3}) were considered and corresponding Eliashberg equations were derived and solved both in the
one-band regime and in the limit of small graphene doping. The Eliashberg function $\alpha^2F$, corresponding to (\ref{V_eff}), is
\begin{eqnarray}
\alpha^2_{\mathbf{v}}(\nu)F(\nu)=\mathcal{N}\sum_\mu g_\mu^2R_\mu(\mathbf{v})\delta(\nu-\omega_\mu),\label{a2F}
\end{eqnarray}
and the equation for the gap in the conduction band $\Delta\equiv\Delta_+(p_\mathrm{F}=\mu/v_\mathrm{F},\omega=0)$ at $T=0$ reads (see
\cite{Lozovik5}):
\begin{eqnarray}
1=2\int\limits_0^{\omega_0}\frac{d\omega}{\sqrt{\omega^2-\Delta^2}}\int\limits_0^\infty
d\nu\frac{\alpha^2_{\mathbf{v}}(\nu)F(\nu)}{\omega+\nu},\label{Sc4}
\end{eqnarray}
where $\omega_0\sim\omega_\mu$ is a cutoff frequency for the gap $\Delta_+(p_\mathrm{F},\omega)$. Solving (\ref{Sc4}) with taking into account
(\ref{a2F}) in the limit $\Delta\ll\omega_0,\omega_\mu$, we obtain the estimate of the gap at $T=0$:
\begin{eqnarray}
\Delta\approx\frac{2\omega_0}{\displaystyle\prod_\mu(1+\omega_0/\omega_\mu)^{\lambda_\mu/\lambda}}\,\exp\left\{-\frac1\lambda\right\},\nonumber
\end{eqnarray}
with the partial $\lambda_\mu=2\mathcal{N}g_\mu^2R_\mu(\mathbf{v})/\omega_\mu$ and total $\lambda=\sum_\mu\lambda_\mu$ coupling constants introduced.

The valley structure of the order parameter $\mathbf{v}$ will adjust itself to achieve the ground state with the lowest energy, corresponding to the
maximal possible value of $\lambda$. When $(g_{\mathbf{K}}^2/\omega_{\mathbf{K}})>2(g_\Gamma^2/\omega_\Gamma)$, the preferable pairing structure is
valley-diagonal: $\mathbf{v}=\{\pi/2,0,0\}$; in the opposite case, when $(g_{\mathbf{K}}^2/\omega_{\mathbf{K}})<2(g_\Gamma^2/\omega_\Gamma)$, the
pairing is valley-off diagonal: $\mathbf{v}=\{0,(\pi/2)\cos\varphi,(\pi/2)\sin\varphi\}$.

Taking the values of the coupling constants from \cite{Piscanec}, we get:
$g_{\mathbf{K}}^2/\omega_{\mathbf{K}}\approx12.02\,\mbox{eV}\cdot\mbox{\AA}^2$, $g_\Gamma^2/\omega_\Gamma\approx3.75\,\mbox{eV}\cdot\mbox{\AA}^2$, so
the valley-diagonal pairing with $\mathbf{v}=\{\pi/2,0,0\}$, when $\Gamma$-phonons compete with the $\mathbf{K}$-phonons, is preferable (note, that
this relation can revert at large dielectric constant of surrounding medium, since $g_{\mathbf{K}}^2$ is highly renormalized to higher values due to
Coulomb interaction \cite{Basko2}). The coupling constant $\lambda\approx0.065\times\mu\,\mbox{[eV]}$ can provide any noticeable pairing only at
heavy chemical doping of graphene with $\mu\approx1.5-2\,\mbox{eV}$. The earlier estimates \cite{Khveshchenko,Lozovik5} of the coupling constants for
electron pairing in graphene, mediated by optical phonons, give similar results.

\section{Electron pairing by out-of-plane phonons}
The consideration of electron pairing by out-of-plane phonons is based on the same approach, as in the previous paragraph, but with using the
interaction Hamiltonian (\ref{H_int_4}) instead of (\ref{H_int_1}). The quadratic electron-phonon coupling results in the loop of two phonon lines
connecting two interacting electrons \cite{Khveshchenko,Mariani1}, instead of one phonon line for linear coupling, as shown in Fig.~\ref{Fig3}(b).
The analogue of self-consistent gap equations (\ref{Sc1}) for the electron-phonon Hamiltonian (\ref{H_int_4}) reads:
\begin{eqnarray}
\Delta_{21}(p)=\frac{2T^2}{S^2}\sum_{p'q}\sum_{\mathbf{Q}\sigma\sigma'}
\overline{V}^{(\mathbf{Q})}_{\mathbf{p}'\mathbf{p}\mathbf{q},\sigma\sigma'}G_{21}(p')
V^{(-\mathbf{Q})}_{\mathbf{p}\mathbf{p}',-\mathbf{q},\sigma\sigma'}\nonumber\\ \times D_\sigma(q)D_{\sigma'}(\mathbf{Q}+p-p'-q),\label{Sc5}
\end{eqnarray}
where
$\overline{V}^{(\mathbf{Q})}_{\mathbf{p}'\mathbf{p}\mathbf{q},\sigma\sigma'}=C^{-1}V^{(\mathbf{Q})T}_{-\mathbf{p}',-\mathbf{p}\mathbf{q},\sigma\sigma'}C$
is the charge-conjugated vertex. The charge conjugation does not change the contributions to the vertices (\ref{vert2}) from bond-stretching, but
changes the sign of the deformation potential contributions. Note, that the summation over $\mathbf{q}$ in (\ref{Sc5}) is performed over the whole
second Brillouin zone.

Performing a summation over $q_0=2\pi i n$ in the phonon loop in (\ref{Sc5}), we get an analogue of phonon Green function, but with the sum of the
frequencies of two phonons:
\begin{eqnarray}
T\sum_{q_0}D_\sigma(q)D_{\sigma'}(\mathbf{Q}+p-p'-q)=\nonumber\\-\frac{2(\omega_{\mathbf{q}\sigma}+
\omega_{\mathbf{Q}+\mathbf{p}-\mathbf{p}'-\mathbf{q},\sigma'})}
{(p_0-p_0')^2-(\omega_{\mathbf{q}\sigma}+\omega_{\mathbf{Q}+\mathbf{p}-\mathbf{p}'-\mathbf{q},\sigma'})^2}.\nonumber
\end{eqnarray}
Further calculations are similar to that in derivation of the Eliashberg equations (\ref{a2F})--(\ref{Sc4}). We consider one-band pairing again,
occurring at $\mu>\omega_{\mathbf{q}\sigma}\sim0.1\,\mbox{eV}$. The Eliashberg function, corresponding to out-of-plane phonons, and entering the
equation analogous to (\ref{Sc4}), is
\begin{eqnarray}
\alpha^2_{\mathbf{v}}(\nu)F(\nu)=-\frac{\mathcal{N}}S\sum_{\mathbf{q}\mathbf{Q}}\sum_{\sigma\sigma'}
\delta(\nu-\omega_{\mathbf{q}\sigma}-\omega_{\mathbf{Q}-\mathbf{q},\sigma'})\nonumber\\
\times\frac14\mathrm{Sp}\left[e^{-i\mathbf{v}\mathbf{T}}\overline{V}^{(\mathbf{Q})}_{\mathbf{q}\sigma\sigma'}\gamma^0e^{i\mathbf{v}\mathbf{T}}\gamma^0
V^{(-\mathbf{Q})}_{-\mathbf{q}\sigma\sigma'}\right].\label{Sc6}
\end{eqnarray}
Here we neglected the momentums $\mathbf{p}$ and $\mathbf{p}'$ in the vertices and phonon frequencies, as in (\ref{vert2}), and performed angle
integration.

It is convenient to rewrite (\ref{Sc4}) in the form:
\begin{eqnarray}
1=\int\limits_\Delta^{\omega_0}\frac{d\omega}{\sqrt{\omega^2-\Delta^2}}\,Z(\omega;\mathbf{v}),\nonumber\\
Z(\omega;\mathbf{v})=\int\limits_0^\infty d\nu\frac{\alpha^2_{\mathbf{v}}(\nu)F(\nu)}{\omega+\nu}.\label{Sc7}
\end{eqnarray}
Summation over $\mathbf{q}$ within the second Brillouin zone in (\ref{Sc6}) with taking into account (\ref{vert2}) results in some effective matrix
operation with the expression $\gamma^0e^{i\mathbf{v}\mathbf{T}}\gamma^0$, enclosed between the vertices. A number of terms under the trace in
(\ref{Sc6}) vanish upon summation, and the remaining nonzero terms allow to write (\ref{Sc7}) as a sum of contributions, corresponding to deformation
potential ($\mu=1,2$) and bond-stretching ($\mu=3,4$) and, on the other side, to the two-phonon processes, leaving the electron in its initial valley
($\mu=1,3$) and flipping it into the opposite valley ($\mu=2,4$): $Z(\omega;\mathbf{v})=\sum_{\mu=1}^4Z_\mu(\omega)R_\mu(\mathbf{v})$. Explicitly, we
have
\begin{eqnarray}
R_1=1,\quad R_2=(-\hat{\mathbf{v}}_2^2+\hat{\mathbf{v}}_3^2)\sin^2v,\nonumber\\
R_3=-\cos^2v-(\hat{\mathbf{v}}_1^2-\hat{\mathbf{v}}_2^2-\hat{\mathbf{v}}_3^2)\sin^2v,\nonumber\\R_4=-\cos^2v+\hat{\mathbf{v}}_1^2\sin^2v,\nonumber
\end{eqnarray}
\begin{eqnarray}
Z_1(\omega)=\frac{\mathcal{N}}S\left(\frac{a^2C_1}{4M}\right)^2\sum_{\mathbf{q}\sigma}\frac{(3-s_\sigma|g_{\mathbf{q}}|)^2}
{(\omega+2\omega_{\mathbf{q}\sigma})\omega^2_{\mathbf{q}\sigma}},\nonumber\\
Z_2(\omega)=\frac{\mathcal{N}}S\left(\frac{a^2C_1}{4M}\right)^2\nonumber\\ \times\sum_{\mathbf{q}\sigma\sigma'}
\frac{(3-s_\sigma|g_{\mathbf{q}}|-s_{\sigma'}|g_{\mathbf{K}+\mathbf{q}}|)^2}
{4(\omega+\omega_{\mathbf{q}\sigma}+\omega_{\mathbf{K}+\mathbf{q},\sigma'})\omega_{\mathbf{q}\sigma}\omega_{\mathbf{K}+\mathbf{q},\sigma'}},\nonumber\\
Z_3(\omega)=\frac{\mathcal{N}}S\left(\frac{a^2C_2}{4M}\right)^2\nonumber\\
\times\sum_{\mathbf{q}\sigma\sigma'}\frac{|s_\sigma
g_{\mathbf{K}+\mathbf{q}}\phi_{-\mathbf{q}}+s_{\sigma'}g_{\mathbf{K}-\mathbf{q}}\phi_{\mathbf{q}}|^2}
{4(\omega+\omega_{\mathbf{q}\sigma}+\omega_{\mathbf{q}\sigma'})\omega_{\mathbf{q}\sigma}\omega_{\mathbf{q}\sigma'}},\nonumber\\
Z_4(\omega)=\frac{\mathcal{N}}S\left(\frac{a^2C_2}{4M}\right)^2\nonumber\\ \times\sum_{\mathbf{q}\sigma\sigma'}\frac{|s_\sigma
g_{-\mathbf{K}+\mathbf{q}}\phi_{-\mathbf{q}}+s_{\sigma'}g_{\mathbf{K}-\mathbf{q}}\phi_{\mathbf{K}+\mathbf{q}}|^2}
{2(\omega+\omega_{\mathbf{q}\sigma}+\omega_{\mathbf{K}+\mathbf{q},\sigma'})\omega_{\mathbf{q}\sigma}\omega_{\mathbf{K}+\mathbf{q},\sigma'}}.\label{Z1}
\end{eqnarray}

\begin{figure}[t]
\begin{center}
\resizebox{0.8\columnwidth}{!}{\includegraphics{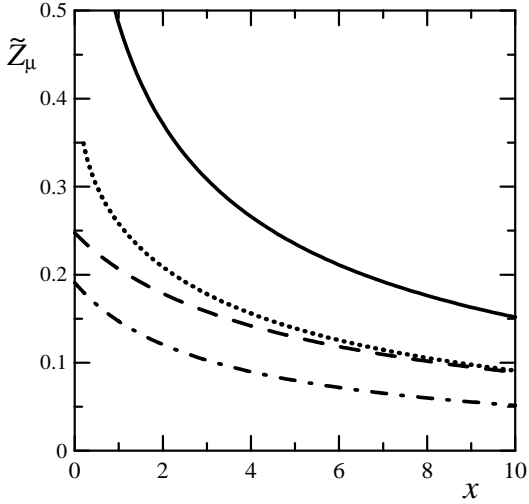}}
\end{center}
\caption{\label{Fig4}The dimensionless functions $\tilde{Z}_\mu(x)$, corresponding to $Z_\mu(\omega)$ (\ref{Z1}). Solid line: $\mu=1$, dashed line:
$\mu=2$, dotted line: $\mu=3$, dash-dotted line: $\mu=4$.}
\end{figure}

The structure of the order parameter, most favorable with respect to all branches $\mu=1,\ldots,4$ jointly, is $\mathbf{v}=\{0,0,\pi/2\}$
(valley-antidiagonal and antisymmetric pairing). In this case, $R_{1,2,3}=1$, $R_4=0$. The functions (\ref{Z1}) can be reduced,
$Z_{1,2}(\omega)=(3\mathcal{N}g_1^2/16a^2M^{1/2}D^{3/2})\tilde{Z}_{1,2}(x)$,
$Z_{3,4}(\omega)=(9\mathcal{N}\gamma^2/32M^{1/2}D^{3/2})\tilde{Z}_{3,4}(x)$, to the dimensionless functions $\tilde{Z}_\mu$ of $x=\omega/\sqrt{D/M}$,
plotted in Fig.~\ref{Fig4}. The functions $\tilde{Z}_1$ and $\tilde{Z}_3$ have logarithmic singularities at $\omega=0$ due to contributions of
acoustical branches, indicating long-range character of electron-electron interaction by out-of-plane phonons (which was noted in
\cite{Khveshchenko}), but generally all $\tilde{Z}_\mu$ are of the order of unity. Therefore, the summary function $Z(\omega)$, playing the role of
the coupling constant in (\ref{Sc7}) is of the order of $10^{-3}$ at $\mathbf{v}=\{0,0,\pi/2\}$, and cannot provide any observable pairing.

Note that consideration of electron pairing in graphene due to out-of-plane phonons, performed in \cite{Khveshchenko}, neglects details of
electron-phonon interaction and includes only acoustical phonon branch, but provides the order of magnitude for the coupling constant, close to that
in our work.

\section{Discussion}
In addition to the previous sections, we put few general remarks. First, the issue of symmetry properties of the order parameter with taking into
account electron spins is worth of discussion. The spin projection indices $s$, $s'$ can be assigned to the anomalous Green functions,
$G^{(21)}_{ss'}(\mathbf{p},\tau)=-\langle T\Psi_{\mathrm{C}\mathbf{p}s}(\tau)\overline\Psi_{\mathbf{p}s'}(0)\rangle$, to give the following condition
of antisymmetry: $G^{(21)}_{ss'}(\mathbf{p},0)=C^{-1}[G^{(21)}_{s's}(-\mathbf{p},0)]^TC$. Applying it to the order parameter of the form
(\ref{Delta_21}), we get that when the pairing is valley-diagonal (i.e. $\mathbf{v}\propto\{1,0,0\}$) or valley-antidiagonal and valley-symmetric
($\mathbf{v}\propto\{0,1,0\}$), the order parameter must have combined spatial-spin symmetry (spin-triplet $s$-wave pairing), unlike the usual
electron-electron pairing \cite{BCS}, which is jointly antisymmetric in the space and spin (spin-singlet $s$- or $d$-wave pairing, or spin-triplet
$p$-wave pairing). Such a peculiarity, as can be shown, is a consequence of additional ``hidden'' antisymmetry of the order parameter by sublattices
(the similar unconventional symmetry of two-electron wave function in graphene was noted in \cite{Sabio}). Conversely, when the pairing is
valley-antidiagonal and valley-antisymmetric ($\mathbf{v}\propto\{0,0,1\}$), the order parameter must be antisymmetric jointly in space and by spins,
as in conventional superconductors.

Another important point is a role of Coulomb repulsion of electrons, which can be added to the self-consistency equations (\ref{Sc1}) with the
vertices $\Gamma^{(1)}=-\Gamma^{(2)}=\gamma^0$. The corresponding angular factor (\ref{ang_fact}) does not depend on $\mathbf{v}$ and always provide
the effective repulsion in the framework of the band-diagonal pairing (\ref{Delta_21}). However, certain forms of the order parameter are possible,
supporting the electron-electron pairing by Coulomb interaction in graphene: for example, the ``vector'' order parameter in \cite{Ohsaku3} or the
resonating valence bond order parameter \cite{Black-Schaffer,Honerkamp}; all of them can be described using the formalism of matrix Green functions
(\ref{G_ij}).

When graphene is heavily doped ($\mu>0.5\,\mbox{eV}$) by impurities (see, e.g., \cite{McChensey}), several additional factors should be taken into
account when considering electron pairing. These are, in particular, an influence of impurities on the condensate \cite{Wehling}, formation of energy
bands of the deposed atoms (similarly to that in graphite intercalation compounds \cite{AlJishi}) and possible structural reconstruction of graphene
\cite{McChensey}. Moreover, the trigonal warping of the Fermi surface in graphene at high doping \cite{CastroNeto} should promote the
valley-antidiagonal pairing of electrons with opposite momenta.

In conclusion, we have considered phonon-mediated electron-electron pairing in graphene taking into account both the details of electron-phonon
interaction resolved by sublattices and valleys, and a possibility of different structures of the order parameter. We assumed an $s$-wave pairing,
diagonal with respect to electron bands, and shown, that conditions of this pairing depend on a structure of the order parameter with respect to the
valley degree of freedom. Contribution of phonon modes in graphene to the effective electron-electron interaction can be attractive, repulsive, or
can even vanish, depending on both the mode symmetry and the valley structure of the order parameter. The orbital-spin part of the order parameter
 can be symmetric in some cases.

We have also considered the quadratic coupling of electrons to out-of-plane phonon modes. Hamiltonian of electron-phonon interaction is derived
taking into account the contributions from deformation potential and from change of bond lengths. Quadratic character of electron-phonon coupling
results in unusual phonon-mediated electron-electron interaction, represented by the loop consisting of two phonon lines. Integration on the inner
momentum in this loop should be performed over the whole Brillouin zone, in contrast to the situation with the linear electron-phonon coupling, where
the phonon momentums are either very small, or connect two electron valleys. The effective action of the out-of-plane phonons on electrons is also
dependent on the structure of the order parameter.

The study of phonon-mediated electron pairing in graphene, presented in this paper, not only allows to extend the analogy between electrons in
graphene and relativistic elementary particles, introducing new kind of interactions in the model (scalar, pseudoscalar, pseudovector phonons etc.),
but would provide better understanding of BCS-like pairing phenomena in unconventional systems.

\section*{Acknowledgement}
The work was supported by the Russian Foundation for Basic Research and by the Program of the Russian Academy of Sciences.

\end{document}